\documentclass[letterpaper,11pt]{article}

\usepackage{jheppub}
\usepackage{graphicx} 
\usepackage{amsmath}
\usepackage{amssymb}
\usepackage{xspace}
 \usepackage{slashed}

\newcommand{\nn}{\nonumber}

\newcommand{\beq}{\begin{equation}}
\newcommand{\eeq}{\end{equation}}
\newcommand{\bea}{\begin{eqnarray}}
\newcommand{\eea}{\end{eqnarray}}

\newcommand{\half}{\frac{1}{2}}
\newcommand{\LN}{\text{ln}}

\def\nbar{\bar n}

\newcommand{\Log}{\text{ln}}
\definecolor{darkred}{rgb}{0.5,0,0}

\newcommand{\binv}{\big(\frac{be^{\gamma_E}}{2}\big)^{-1}}

\allowdisplaybreaks[2]
\preprint{MIT--CTP 4646}
\title{The Higgs Transverse Momentum Distribution at NNLL and its Theoretical Errors}

\author[a]{Duff Neill,}
\author[b]{Ira Z. Rothstein}
\author[b]{Varun Vaidya}
\affiliation[b]{Department of Physics, Carnegie Mellon University, Pittsburgh, PA~15213, U.S.A.}
\affiliation[a]{Center for Theoretical Physics, Massachusettes Institute of Technology, Cambridge MA~ 022139  U.S.A.}

\abstract{
In this letter, we present the NNLL-NNLO transverse momentum Higgs distribution arising
from gluon fusion. In the regime $p_\perp\ll m_H$ we include the
resummation of the large logs at next to next-to leading order and then match on
to the $\alpha_s^2$ fixed order result near $p_\perp \sim m_h$.  
By utilizing the rapidity renormalization group (RRG) we are able to smoothly match between
the resummed, small $p_\perp$ regime and the fixed order regime.
We give a detailed discussion of
the scale dependence of the result including an analysis of the rapidity scale
dependence.
Our central value differs from previous results,  in the transition region as well as  the tail, by an amount which is outside the error band. This difference is  due to the fact that the
RRG profile allows us to smoothly turn off the resummation.  }
\begin{document}

\maketitle
\section{Introduction}

We are  transitioning into an era of precisions Higgs physics. Presently there is sufficient data to study various decay modes of the Higgs \cite{decays}, and soon there will  be sufficient data to study the  the Higgs differnential cross section, which can be used to better understand the underlying production mechanism and to search for new physics \cite{NP}. While the new physics would be most prominent at large values of  $p_\perp$, the predominance of the events will be in the lower range. Furthermore, to isolate the Higgs' decays from backgrounds, events are binned according to their highest-transverse momentum jet. The 0-jet bin, corresponding to no central jets above a certain transverse-momentum threshold, plays a critical role in the current Higgs analysis \footnote{Resummed predictions for the 0-jet bin  can be found  in \cite{banfi_pt_veto,becher_pt_veto_1, becher_pt_veto_2, frank_pt_veto_1, frank_pt_veto_2}.}. Thus there is increased motivation for making precise predictions for the distribution when $p_\perp\ll m_h$  since back to back central jet production is power suppressed \cite{RRG2}.

 The small $p_\perp$  region of parameter space is polluted by large logarithms which must be resummed in order to retain  systematic control of the theoretical errors.  Resummations have been previously been discussed within an  SCET \cite{SCET}  framework [\cite{mantry_higgs},\cite{drell_becher},\cite{higgs_becher},\cite{mantry_higgs_II},\cite{Gao_higgs},\cite{idilbi_higgs},\cite{idilbi_TMDs}, \cite{idilbi_pt_NNLL}, \cite{idilbi_QCD_np_pt_distr}] as well as in the CSS resummation formalism [\cite{CSS},\cite{Nova_higgs},\cite{kaufmann},\cite{Florian},\cite{catani},\cite{Grazzini_I}].
While formally most of the these results agree at a given order in the resummation procedure, the central values as well as the predicted errors between different resummation techniques will differ\footnote{For a detailed comparison of resummation methods in the context of $e^+e^-$ event shapes see \cite{sterman_qcdvscet}, and for threshold resummations see \cite{Bonvini_Threshold, Sterman_Threshold}.}. Another crucial difference between various theoretical predictions is how the transition between the fixed order result at large $p_\perp$ and the resummed result is handled. In this paper we  present a result for the resummed cross section at NNLL + NNLO 
within the confines of SCET and the rapidity renormalization group (RRG) \cite{RRG1,RRG2}. 
Our motivation for this analysis is two-fold.
By working within the RRG formalism we are able to consistently turn off the resummation
in the tail region.  It has been shown in the context of $B\rightarrow X_s +\gamma$ \cite{bsg}, thrust \cite{thrust} as well
as the Higgs jet veto calculation \cite{BS}, that if one does not turn off the resummation then one can over estimate the cross section, in the region where fixed order perturbation theory should suffice,  by an amount which goes beyond the canonical error band in the fixed order result.
This overshoot happens despite the fact that the resummed terms are formally sub-leading
in the expansion.  The reason for this overshoot has  been shown \cite{bsg,thrust,BS}  to be due to the fact that
there are cancellations between the singular and non-singular terms in the tail region and that
this cancellation will occur only if the proper scale is chosen in the logarithms.  This will be
born out in our analysis as well. 
By using the RRG we have a natural way to interpolate between the resummed 
and fixed order regions.
This allows us to present the full spectrum such that in the large $p_\perp$ region
our results match onto the NNLO result, as previously discussed within the context of
the jet veto cross section \cite{frank_pt_veto_1,frank_pt_veto_2}. In this sense our paper fills a gap in the literature.
As we will discuss below, the RRG plays a important role in the theory error determination.  We also discuss in detail how our error estimates and full spectrum differ from those of previous work.


\section{Systematics}

In this section, for the sake of completeness,  we review the well known systematics of the calculation.  We work within the confines of the large top mass approximation. This approximation is known to work extremely well - much better then one would naively expected - especially away from the tail of the spectrum. We will be not be considering $p_\perp$ large enough for these corrections to be relevant to the error budget. A full discussion of such errors as well as others will be discussed at the end of the paper. 

 At the scale $m_t$, full QCD is  matched onto the large top mass effective theory at order $\alpha_s^2$, after which  we match onto SCET at the scale $m_h$. In SCET we will work to leading order in a systematic expansion in $p_\perp/m_h$, and the errors due to non-perturbative corrections, which are suppressed by $\Lambda/p_\perp$, are included in the final error analysis. 

As is well known, logs in the perturbative expansion - in impact parameter space - exponentiate and allow us to organize the series as follows
\beq
\sigma \sim \text{Exp}\Bigg[\text{ln}\Big(bm_h \Big)F\Big(\alpha_s \text{ln}\Big(b m_h\Big)\Big) +G\Big(\alpha_s \text{ln}\Big(bm_h\Big)\Big)+\alpha_s H\Big(\alpha_s \text{ln}\Big(bm_h\Big)\Big) +....\Bigg].
\eeq 
The standard terminology is such that keeping only $F$ corresponds to ``LL''   leading (double) log, while if $G$ is retained we would have NLL etc. As $p_\perp$ gets larger the EFT breaks down and the fixed order calculation becomes the relevant quantity.  We will utilize the order $\alpha^2$ (NNLO) result \cite{Ravindran:2002dc}, \cite{glosser}\footnote{As is common with fixed order calculations, the authors of \cite{glosser} use the term NLO, since the leading order result is a delta function in $p_\perp$.} in this large $p_\perp$ region. We do so despite the fact in the small $p_\perp$ region we will be only keeping terms of order $\alpha$. This is not inconsistent as the error in the two disparate regions are distinct and uncorrelated. In the resummed region we will be working at NNLL in the perturbative expansion and at leading order in the power expansion $p_\perp/m_h$. By utilizing the RRG scale we will smoothly turn off the resummation and match onto the full NNLO fixed order calculation.

\section{Factorization in SCET and anomalous dimension}

A factorization theorem for Higgs production at small $p_\perp$  within SCET and the RRG formalism was developed in \cite{RRG2}.
The starting point in the large top mass effective theory is the gluon fusion operator 
\begin{equation}{\cal H}(x) = C_t\frac{h(x)}{v} {\rm Tr} [G^{\mu\nu}(x)G_{\mu\nu}(x)].
\end{equation}
The matching coefficient for this operator ($C_t$)  is known to two loops \cite{Inami}.

At the scale $m_h$ \footnote{We ignore the running between the top mass scale and $m_h$ as these logs will be sub-leading in our power counting.}  we match onto SCETII. This theory is the version of SCET in which the collinear modes and the soft modes have the same invariant mass, which distinguishes it from SCETI where there exists a hierarchy in masses between these two modes.  Due to this equality in virtualities a new set of divergences - rapidity divergences - arise, which are not regulated by dimensional regularization. The rapidity regulator introduces a new scale which acts as a boundary between the collinear and soft modes as discussed in \cite{RRG1,RRG2}. The reader may consult (\cite{RRG2}) for the  details of the formalism. At leading order in $\lambda=p_\perp/M_h$ the differential cross section for higgs production at low transverse momentum may be written as
\begin{multline}
\label{fact}
\frac{d\sigma}{d p_{\perp}^2 dy}= \frac{C_t^2}{8v^2S(N_c^2-1)}\int \frac{d^4p_h}{(2\pi)^4}(2\pi)\delta^{+}(p_h^2-m_h^2) \delta\!\!\left(y-\half \ln\frac{p_h^+}{p_h^-}\right) \delta(p_{\perp}^2 - \vert \vec p_{h\perp}\vert^2)\\
4(2\pi)^8\int d^4x e^{-i x\cdot p_h}H(m_h)f_{\perp\, g/P}^{\mu\nu}(0,x^+,\vec{x}_{\perp})f_{\perp\, g/P\,\mu\nu}(x^-,0,\vec{x}_{\perp}){\cal S}(0,0,\vec{x}_{\perp})
\end{multline}
$\cal S$ is the soft function defined as 
\begin{align}
\label{soft}
\mathcal{S}(0,0,\vec{x}_{\perp})&=\frac{1}{(2\pi)^2(N_c^2-1)} \langle0|S_{n}^{ac}(x)S_{\bar n}^{ad}(x)S_{n}^{bc}(0)S_{\bar n}^{bd}(0)|0\rangle \, , \nn\\
\end{align}
defined in terms of the light-like Wilson lines
\beq
S_{n}(x)= P \exp \left(ig\int_{-\infty}^0 n \cdot A_s(n \lambda+x)d\lambda \right).
\eeq
 $f^{\alpha \beta}$ is the transverse momentum distribution function (TMPDF) which is matched onto the PDFs at the scale $p_\perp\gg\Lambda$.
\begin{align}
f_{\perp\, g/P}^{\mu\nu}(0,x^+,\vec{x}_{\perp})&=\frac{1}{2(2\pi)^3}\langle p_n|[B_{n\perp}^{A\mu}(x^{+},\vec{x}_{\perp})B_{n\perp}^{A\nu}(0)]|p_n\rangle \, , \\
f_{\perp\, g/P}^{\mu\nu}(x^-,0,\vec{x}_{\perp})&=\frac{1}{2(2\pi)^3}\langle p_{\bar n}|[B_{\bar n\perp}^{A \mu}(x^{-},\vec{x}_{\perp})B_{\bar n\perp}^{A\nu}(0)]|p_{\bar n}\rangle \nn
\end{align}
defined in terms of 
\beq
B_{n\perp}^{a\mu}(x)=\frac{2}{g}{\rm Tr} \left[ T^a \left[ W_n^\dagger(x)i { D}^\mu_{n\perp}W_n(x)\right] \right].
\eeq
$W_n$ is a collinear Wilson line in the fundamental representation defined in $x$-space by
\beq
W_n(x)=P \exp \left(ig\int_{-\infty}^x \bar n \cdot A_n(\bar n \lambda)d\lambda \right).
\eeq
$n/\bar{n}=(1,0,0,\pm1)$ are the null vectors  which correspond to the directions  of the incoming protons.
In impact parameter space, we can write the functions $S$ and $f^{\alpha \beta}$ in terms of their inverse Fourier transform.
\begin{eqnarray}
S(b)= \int{\frac{d^{2}\vec{P_{s\perp}}}{(2\pi)^{2}}}e^{i\vec{b}.\vec{P_{s\perp}}}S(\vec{P_{s\perp}})\\
f_{n}^{\mu\nu}(b,z)= \int{\frac{d^{2}\vec{P_{\perp}}}{(2\pi)^{2}}}e^{i\vec{b}.\vec{P_{\perp}}}f_{n}^{\mu\nu}(\vec{P_{\perp}},z).
\end{eqnarray}
The cross section is given by
\begin{equation}
\frac{d^{2}\sigma}{dP_{t}^{2}dy}=\pi(2\pi)^{5}\frac{H(M_h) \pi C_{t}^{2}}{2v^{2}s^{2}(N_{c}^2-1)}\int{dbbJ_{0}(bP_{t})f_{n}^{\alpha\beta}(b,z_{1})f_{\alpha\beta,n'}(b,z_{2})S(b)}
\end{equation}
$H(M_h)$ is the hard coefficient which depends on the scale $\mu=M_h$ and  and  we define $H(M_h)=8M_h^2|C_s|^2$, where the coefficient $C_s$ is known to two loops \cite{harlander}. The soft function as well as the TMPDFs depend upon both $\mu$, the usual scale introduced within the context of dimensional regularization which factorizes the hard modes from the soft and collinear modes, as well as $\nu$,  the rapidity factorization scale which distinguishes between the soft and collinear modes.
\section{Renormalization and Resummation}\label{sec:Renorm_Resum}
Each of the pieces of the factorization theorem $H$, $\cal S$ and $f$, have natural scales with which they are associated.  $H$ is independent of $\nu$ and its natural $\mu$ scale is $m_h$. The soft and collinear functions have natural scales $(\mu=p_\perp,\nu=p_\perp)$ and $(\mu=p_\perp,\nu=m_h)$ respectively. When these objects are evaluated at these scales, they will be devoid of large logarithms. However, given the natural distribution of scales we can see that it is not possible to choose a $\mu$ and $\nu$ such that all of the individual pieces sit at their respective natural scales. Thus it is expedient to choose a $(\mu,\nu)$ value such that the maximum number of pieces sit at their natural scale, and then use the RG and RRG to sum logs for the  remaining pieces.

 The individual functions  satisfy the following RGE's.
\begin{eqnarray}
\mu \frac{d}{d\mu}H(\mu) = \gamma_{\mu}^{H} H(\mu)\nn \\ 
\kappa_i\frac{d}{d\kappa_i}F(b,\mu,\nu) = \gamma_{\kappa_i}^{F} F(b,\mu,\nu)  \\  
\end{eqnarray}
where $\kappa_i=(\mu,\nu)$ and $F \in \{{\cal{S}},f^{\alpha \beta}\}$. The anomalous dimensions satisfy the relations 
\begin{eqnarray}
\gamma_{\mu}^{H}+\gamma_{\mu}^{S}+2\gamma_{\mu}^{f}=0\nn \\
\gamma_{\nu}^{S}+2\gamma_{\nu}^{f}=0.
\end{eqnarray}
All the renormalization group equations are diagonal in impact parameter space and hence are straightforward to solve. The scales for RG and RRG operations commute, and as a consequence of this, the following relations are generated
\begin{eqnarray}
\mu\frac{d}{d\mu}\gamma_{\nu}^{S}=\nu\frac{d}{d\nu}\gamma_{\mu}^{S}= -2\Gamma_{cusp}\\
\mu\frac{d}{d\mu}\gamma_{\nu}^{f}=\nu\frac{d}{d\nu}\gamma_{\mu}^{f}= \Gamma_{cusp},
\end{eqnarray}
where $\Gamma_{cusp}$ is the cusp anomalous dimensions of two light-like Wilson lines \cite{korchemsky}.

\begin{figure}
\centering
\includegraphics[width=4in]{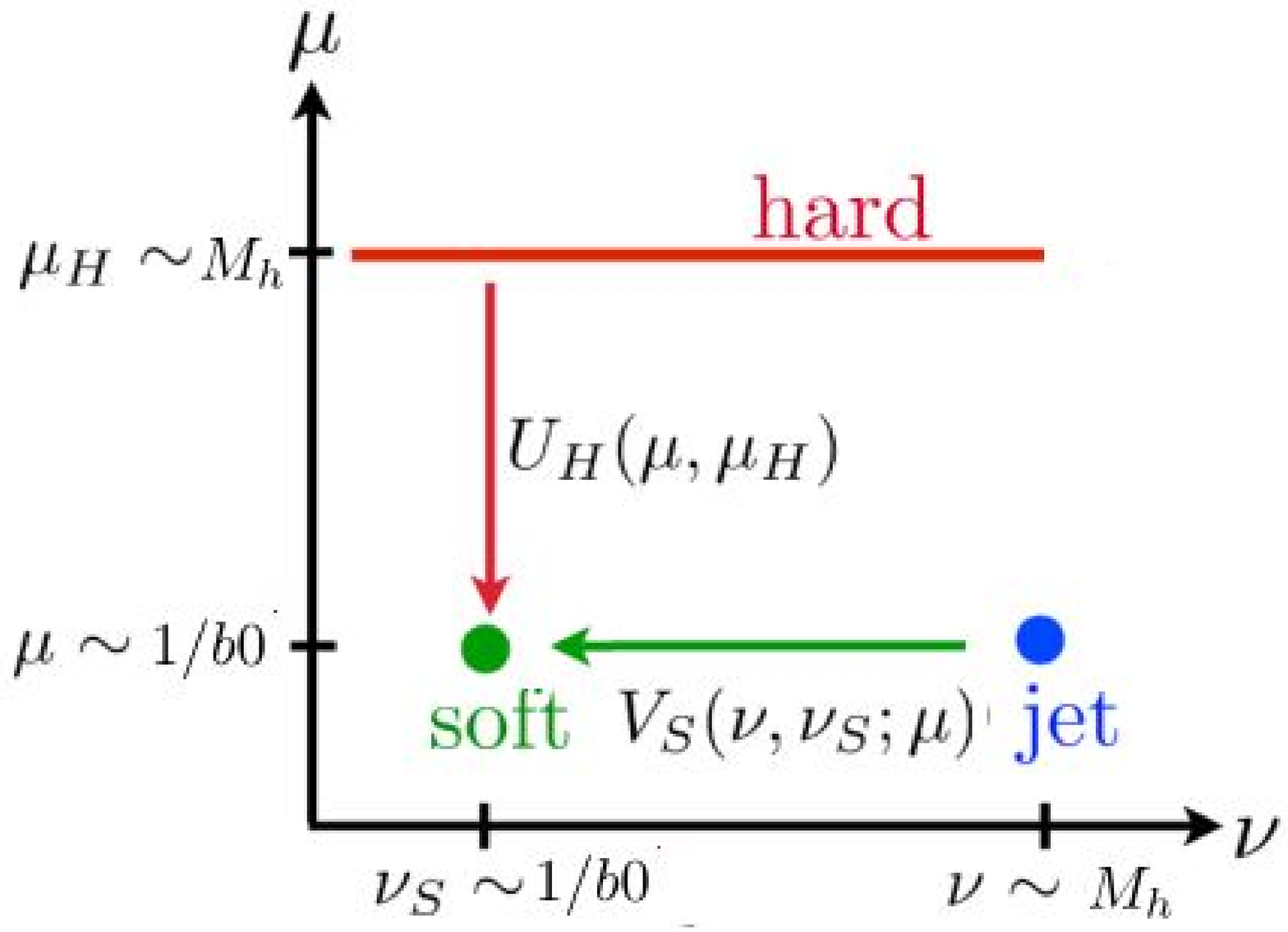}
\caption{The path in $(\mu,\nu)$ space used to resum the logs.}
\label{resum}
\end{figure}

Figure (\ref{resum}) shows the path in $(\mu,\nu)$ space we have chosen to resum the logs.
The hard part is run from the scales $m_h$ down to the scale $1/b_0\sim p_\perp$, while
the jet function is run in $\nu$ space up to the scale $m_h$, where the scale $b_0$ is defined as $b e^{\gamma_E}/2$.
In performing the hard resummation,  $\pi^2$ are resummed by analytically continuing the matching scale from space-like to time-like kinematics relevant for Higgs production\cite{Magnea:1990zb}. It has been shown that  the inclusion of the $\pi^2$ in the resummation does improve the perturbative convergence of the hard matching expansion \cite{Ahrens:2008nc, Ahrens2}, however this method only sums a subset of $\pi^2$  and one can not
claim that this method leads to a systematic reduction in theory errors \footnote{There is no scaling rule for these terms}. 


At NNLL, the systematics require that we keep the cusp anomalous dimensions at three loops, the non-cusp at two loops and matching at both the high and low scales at order $\alpha$.
All of the necessary anomalous dimensions have been calculated previously in the literature. The hard anomalous dimensions can be extracted from  \cite{2loopnoncusp}. We utilize  the one loop matching for the jet and soft function from (\cite{RRG2}).  To run the jet function in $\nu$ we need only the two loop cusp piece since  log in jet rapidity the anomalous dimensionvis not large.  The coefficient of this log is fixed by the hard function anomalous dimensions. We also the need the two loop ``non-cusp'' piece in $\nu$ which can be obtained from the two loop jet function.  The results for the hard anomalous dimension are tabulated in \cite{BS}.
The matching of the TMPDF onto the PDF may be written as 
\begin{multline}
\label{TMDPDF_PDF_matching}f_{\perp g/P}^{R\,\mu\nu}(z,\vec{p}_{\perp})=\sum_{k}\frac{1}{z}\int_{z}^{1}\frac{dz'}{z'}\Big\{\frac{g_{\perp}^{\mu\nu}}{2}I_{\perp 1\, g/k}(z/z',\vec{p}_{\perp}^2)\\
+\Big(\frac{\vec{p}_{\perp}^{\,\mu}\vec{p}_{\perp}^{\,\nu}}{\vec{p}_{\perp}^{\,2}}+\frac{g_{\perp}^{\mu\nu}}{2}\Big)I_{\perp 2\, g/k}(z/z',\vec{p}_{\perp}^2)\Big\}f_{k/P}^{R}(z')+O\Big(\frac{\Lambda_{QCD}}{|\vec{p}_{\perp}|}\Big),
\end{multline}
where the sum is on species of partons, and the parton PDF's are defined as
\bea
f_{g/P}(z)=  - z\,  \bar n \cdot p_n \theta(z) \, g_{\perp\mu\nu}\langle p_n\mid
\left[ B^{c \mu}_{n \perp}(0)
\delta(z \, \bar n \cdot p_n -{ \cal {\bar P}} ) B_{n \perp}^{c \nu}(0)\right] \mid p_n \rangle. \nn\\
f_{q/P}(z)=  - z\,  \bar n \cdot p_n \theta(z) \, \langle p_n\mid
\left[\bar \xi^q_{n,p}(0) W_n \frac{ \slashed {\bar n}}{2}
\delta(z \, \bar n \cdot p_n -{ \cal {\bar P}} )W_n^{\dagger}\xi^q_{n,p}(0) \right] \mid p_n \rangle.
\eea
We adopt the mostly minus metric such that conventions that $\vec{p}_{\perp}^{\,\alpha} \vec{p}_{\perp}^{\,\beta} g_{\perp\alpha\beta}=-\vec{p}_{\perp}^{\,2}$ and  make use of the 't Hooft-Veltmann scheme, so the external transverse momenta remains in 2 dimensions, as do the external polarizations on the operator (the free Lorentz induces). The scheme choice is advantageous, as it allows one to renormalize the operator directly in $\vec{p}_{\perp}$ space. At tree level  the TMDPDF matches on to the PDF as follows:
\begin{align}
\label{TMDPDFp_tree}f_{\perp\, g/g}^{(0)\alpha\beta}(z,\vec{p}_{\perp})&=\delta(1-z)\delta^{(2)}(\vec{p}_{\perp})\frac{g^{\alpha\beta}_{\perp}}{2}f_{g/P}(z).\\
\end{align}
 The one loop renormalized TMPDF, necessary  to extract the order $\alpha$ matching coefficients ($I^{(1,2)}_{\perp ig/g}$), is given by
 \begin{align}
I_{\perp 1\, g/g}(z/z',\vec{p}_{\perp}^2)&=\frac{\alpha_sC_A}{\pi}{\cal L}_0\Big(\mu,\frac{\vec p_\perp}{\mu}\Big)\Bigg(-\ln\Big(\frac{\nu^2}{\omega_-^2}\Big)\delta(1-z)+p_{gg*}(z)\Bigg)\nonumber\\
I_{\perp 2\, g/g}(z/z',\vec{p}_{\perp}^2)=&-2\frac{\alpha_sC_A}{\pi}\frac{1-z}{z} {\cal L}_0\Big(\mu,\frac{\vec p_\perp}{\mu}\Big)\left(\frac{\vec{p}_{\perp}^{\,\alpha} \vec{p}_{\perp}^{\,\beta}}{\vec{p}_{\perp}^{\,2}}+\frac{g_{\perp}^{\alpha\beta}}{2}\right) \,\nn\\
I_{\perp 1\, g/q}(z/z',\vec{p}_{\perp}^2)=& \frac{\alpha_sC_F}{\pi}\Bigg({\cal L}_0\Big(\mu,\frac{\vec p_\perp}{\mu}\Big)P_{gq}(z) + \delta^{(2)}(\vec{p}_{\perp})R_{gq}(z)\Bigg)\nonumber\\
\end{align}
where we have written the expression in terms of plus distribution ${\cal L}_n=\frac{1}{2\pi\mu^2}\Big[\frac{\mu^2}{\vec{p}_\perp^{\,2}}\ln^n\Big(\frac{\mu^2}{\vec{p_\perp}^2}\Big)\Big]_{+}^{1}$, the properties of this distribution are collected in the appendix of \cite{RRG2}.
 
The resummation is carried out in impact parameter space and when Fourier transforming back to $p_\perp$ space, the integrand diverges along the path of integration due to the Landau pole at $1/b \sim \Lambda_{QCD}$. However looking at the form of the cross section, the integrand over $b$ contains the Bessel function of first kind with argument $bp_\perp$ and an exponent which contains powers of $\sim log(M_hb)$. The combination of these two factors quickly damps out the integrand outside the region $1/p_\perp>b>1/m_h$. Thus to avoid hitting the pole, we can simply put a sharp cut off for $b$ at a value sufficiently higher than the largest value of $1/p_\perp \sim 0.5 GeV^{-1}$  but smaller than $1/\Lambda_{QCD}$. A convenient value is found to be $b\sim 2 $GeV$^{-1}$. It has been numerically ascertained that the variation of this cut-off between the values of $1.5-3~ GeV^{-1}$ produces an error only in the fourth significant digit of the integral. 
\section{ The Fixed Order Cross Section and Power Corrections}
When we match the full theory onto SCET we drop terms which are beyond leading order in $\lambda=p_\perp/M_h$. However, a comparison of the full theory fixed order cross section(\cite{glosser}) with the effective theory calculation reveals that in order to achieve ten percent accuracy, we need to include these higher order terms in $\lambda$ beyond $p_\perp> 30$ GeV. In principle we could achieve this by keeping higher order terms in the factorization theorem. However, this would only be necessary if we wished to resum  the logs associated with these power
 corrections. Given that these power corrections are only relevant in regions of larger $p_\perp$, 
 these logs are not numerically large and thus represent a small corrections which need not be resummed. At large $p_\perp$ we should turn off the resummation and use the
 fixed order NNLO result. Thus we subtract out the logs from the NNLO correction so that
 we do not double count, and turn off the resummation using scale profiles which we now consider.

\section{Profiles}
The SCET formalism we have used retains the leading order operators in the power counting parameter $\lambda$. This restricts the validity of the resummed cross-section to the regime where $p_\perp<<M_h$.  At larger values of of $p_\perp/M_h$ the resummation becomes superfluous.  Moreover, the power corrections from subleading operators become increasingly relevant and to maintain accuracy it is vital to include these power corrections, as discussed above. So it is desirable to smoothly switch over from the resummed result to the full theory fixed order cross section when the singular and non-singular terms of the cross-section are comparable.
Notice that formally, it is not necessary to turn off the resummation. As long as one makes the
proper subtraction to avoid double counting, then the resummed terms that are kept should
be sub-leading in the logarithmic power counting, since we are in the transition region. However, as discussed in the introduction, the fixed order
result involves the cancellation between singular and non singular terms. If the scale of the log is
not chosen appropriately (of order $p_\perp$) then this cancellation is not effective and the cross section
is over estimated beyond the canonical scale variation errors \footnote{We thank Iain Stewart for discussions on this point.}. This will be demonstrated below.

\begin{figure}
\centering
\includegraphics[width=4in]{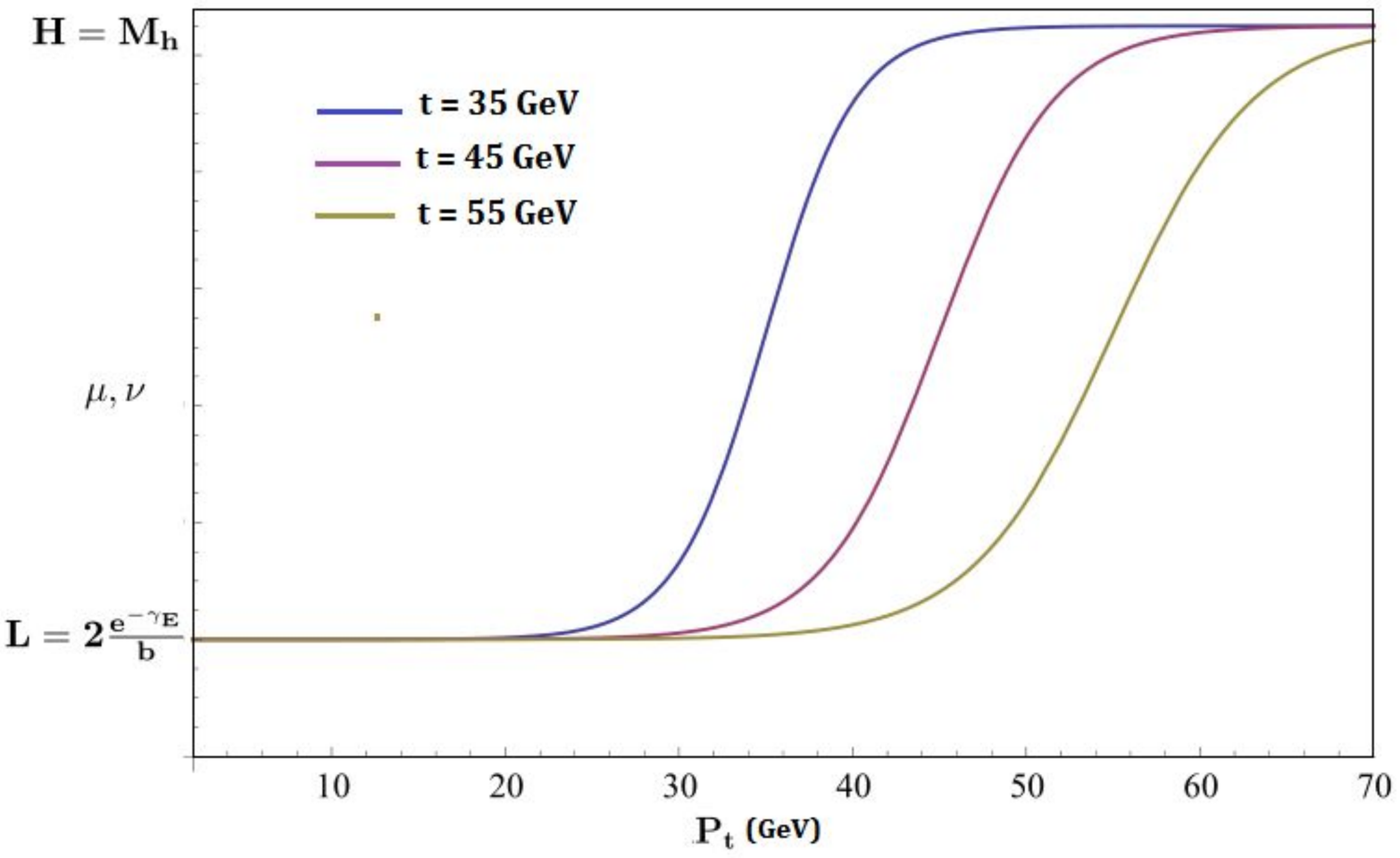}
\caption{Profiles for turning off resummation}
\label{pro}
\end{figure}

The shutting down of the resummation is achieved by varying the low matching scale from its value $1/b_0$ in the low $p_\perp$ region to  $M_h$ in the high $p_\perp$ region. This has the effect of turning off the resummed exponent. To do this, we introduce profiles in $\mu,\nu$ following the work done in \cite{frank_pt_veto_1,frank_pt_veto_2}.   
We use three typical profiles for varying the renormalization scale from $1/b_0$ to $M_h$. Each profile is chosen as a linear combination of hyperbolic tangent functions which smoothly transitions from the resummation region to the fixed order one. The general equation for each profile $P$ can be written as 
\begin{equation}
P(s,L,H,P,t)= \frac{L}{2}\Bigg(1-\text{tanh}\Big[s\Big(\frac{4P}{t}-4\Big)\Big]\Bigg)+ \frac{H}{2}\Bigg(1+\text{tanh}\Big[s\Big(\frac{4P}{t}-4\Big)\Big]\Bigg),
\end{equation}
where $s$ determines the rate of transition,  $L=2e^{-\gamma_{E}}/b$ is the initial value  and $H=M_h$ is the final value. $t$ is the value about which the transition is centered. Fig. (\ref{pro}) shows three profiles with $s=2$  and  $t=(35,45,55)~GeV$.

%


\section{Error analysis}
In our formalism, we have performed an expansion of the cross-section in  $M_h/m_t$, $p_\perp/M_h$, $\Lambda_{QCD}/p_\perp$ and the strong coupling $\alpha_s$. Therefore, the dominant error in our cross-section is due to the first sub-leading term that we drop in each of these parameters.
When expanding in $M_h/m_t$, we retain only the leading order operator. A comparison of the cross section in this limit with  the exact $m_t$ dependent result  \cite{glosser,largemt}, reveals that this approximation works extremely well for $p_\perp < m_t$.  For the range of  $p_\perp$ discussed in this paper, the error due to these corrections  is less than 1 $\%$.
For the parameter $p_\perp/M_h$ , again we retain only the leading order operator. This approximation works well for $p_\perp \leq$ 30 GeV as the leading 
correction scales as $p_\perp^2/M_h^2$. As discussed earlier, for larger values of $p_\perp$, we include the full NNLO expression which includes all orders in $p_\perp/M_h$. 

For the third parameter, $\Lambda_{QCD}/p_\perp$ one must consider the non-perturbative contributions.
Unlike the case of a central veto \cite{BS} there is more then one relevant non-perturbative parameter.
The reason is that we are working in SCETII where the collinear and soft function both have the same invariant mass scale. As such, there will be non-perturbative contributions coming from both of these sectors. As $p_\perp$ approaches  $\Lambda_{QCD}$, we need to systematically include higher dimensional operators in both the soft and the beam sectors. There are multiple non-perturbative functions which contribute in this regime.
In principle these matrix elements can be fit to the data. 
Work in this direction has been performed for vector boson production where ans{\"a}tzes \cite{davies} for
these matrix elements have been  fit to the data \cite{np}.
However, since we are interested in a gluon initiated process we can not hope to use any
extraction from the quark initiated vector boson production process.
Therefore we make no attempt to model these corrections and simply include a
 rough estimate of the errors due to these terms by assuming an error which scales as  $ \Lambda^2_{QCD}/p_\perp^2$.

Finally we consider the errors due to higher order perturbative corrections.
The accepted methodology for estimation of these  errors is to  vary the scale $\mu$ by factors 
of two. The idea is that such variations estimate the size of the constant terms which
are not captured by the renormalization group, as well as the sensitivity to the perturbative trunctation of the anomalous dimensions.  Of course this is just a rough guess which
falls into the rubric of ``you do the best you can," though it is important to try to gauge all sources of large logarithms in these variations.

When working with the RRG we generalize this technique since we now have two distinct types of
factorizations, virtuality and rapidity, and  within the  RRG there are therefore  two 
distinct exponentiations.
In both cases there is a choice as to what sub-leading terms should go into
the exponent. 
Varying the scales $\mu,\nu$ corresponds to varying the size of these sub-leading pieces.  Given that virtuality and rapidity factorization are independent mechanisms
we should  vary both of these scales in order to estimate the errors in the choice
of exponents. 
\begin{figure}
      \centering
			\includegraphics[width=4in]{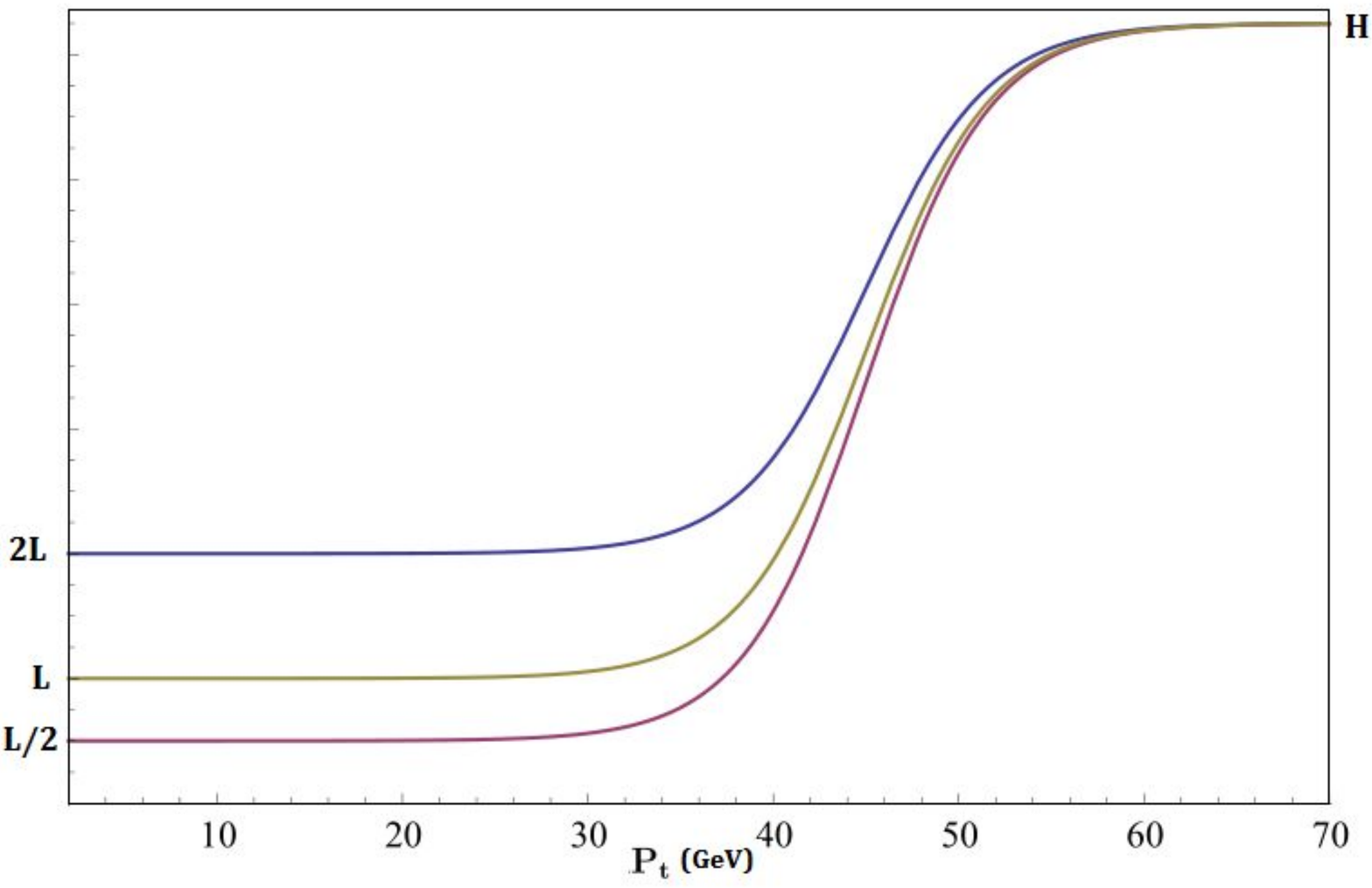}
    \caption{low $p_T$ scale variation}
    \label{var1}
  \end{figure}
	
	\begin{figure}
      \centering
			\includegraphics[width=4in]{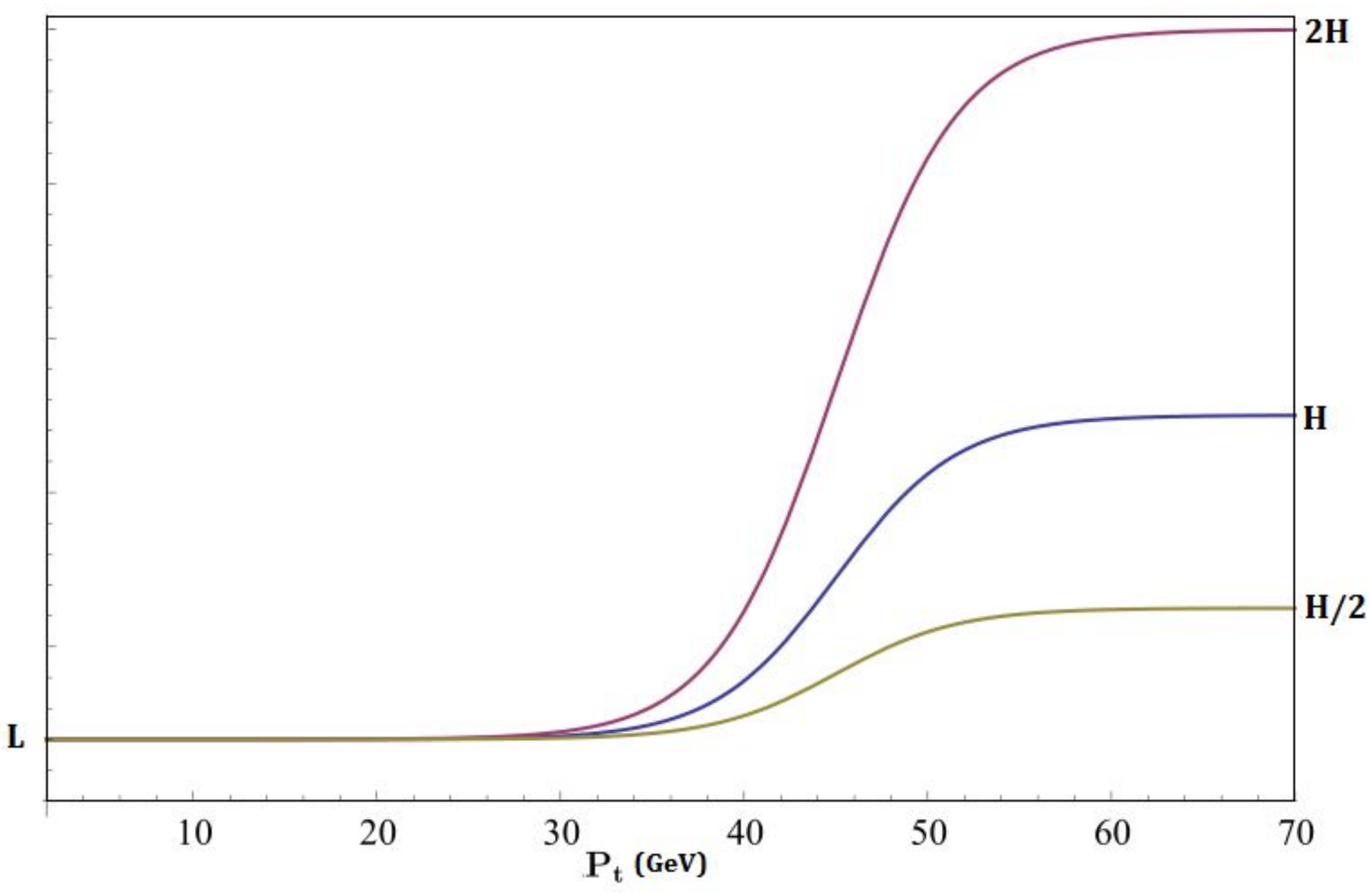}
    \caption{High $p_T$ scale variation}
    \label{var2}
  \end{figure}

%

\begin{figure}
      \centering
			\includegraphics[width=5in]{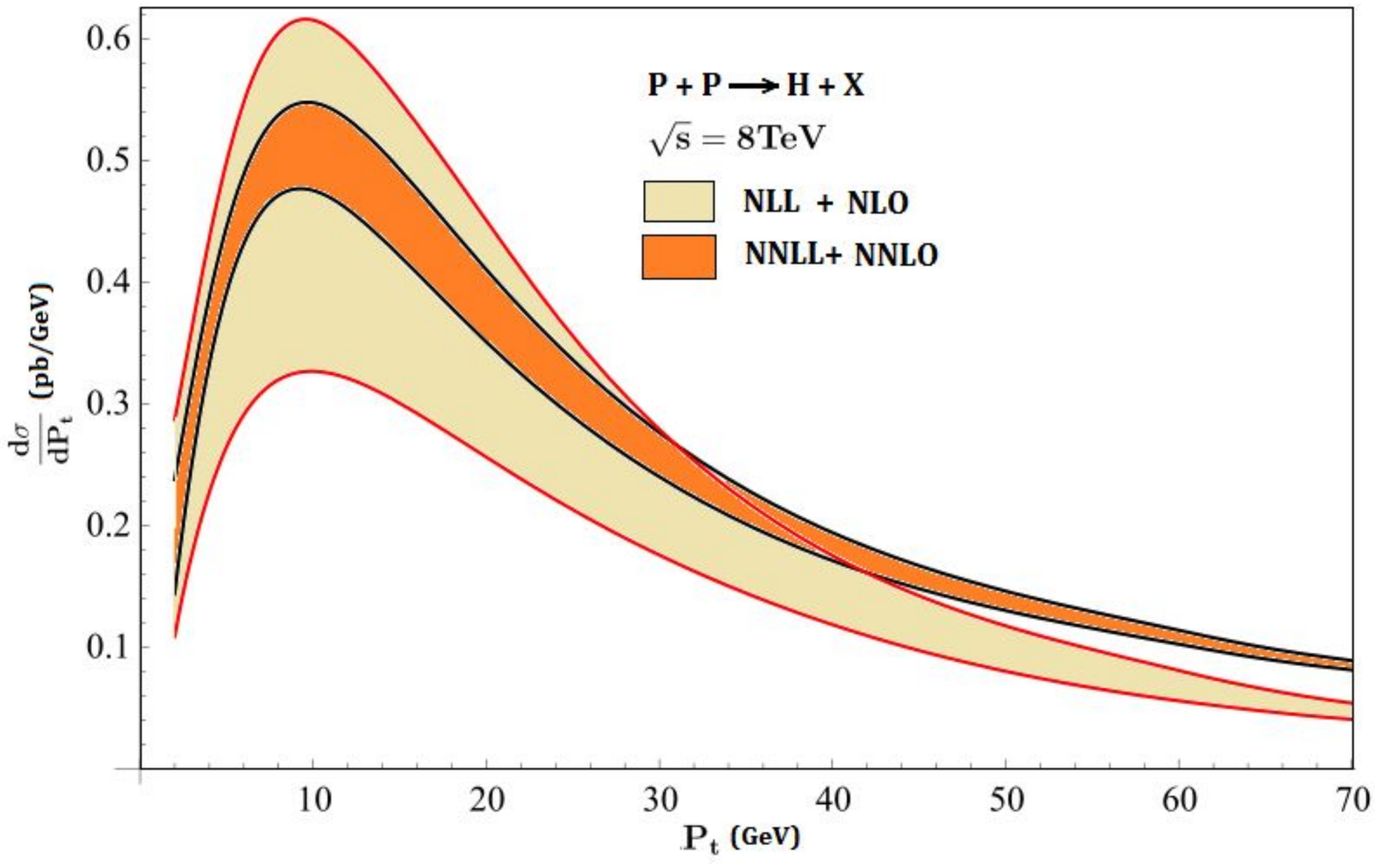}
    \caption{Cross section at 8 TeV}
    \label{8TeV}
  \end{figure}
\begin{figure}[b]
	\centering
    \includegraphics[width=5in]{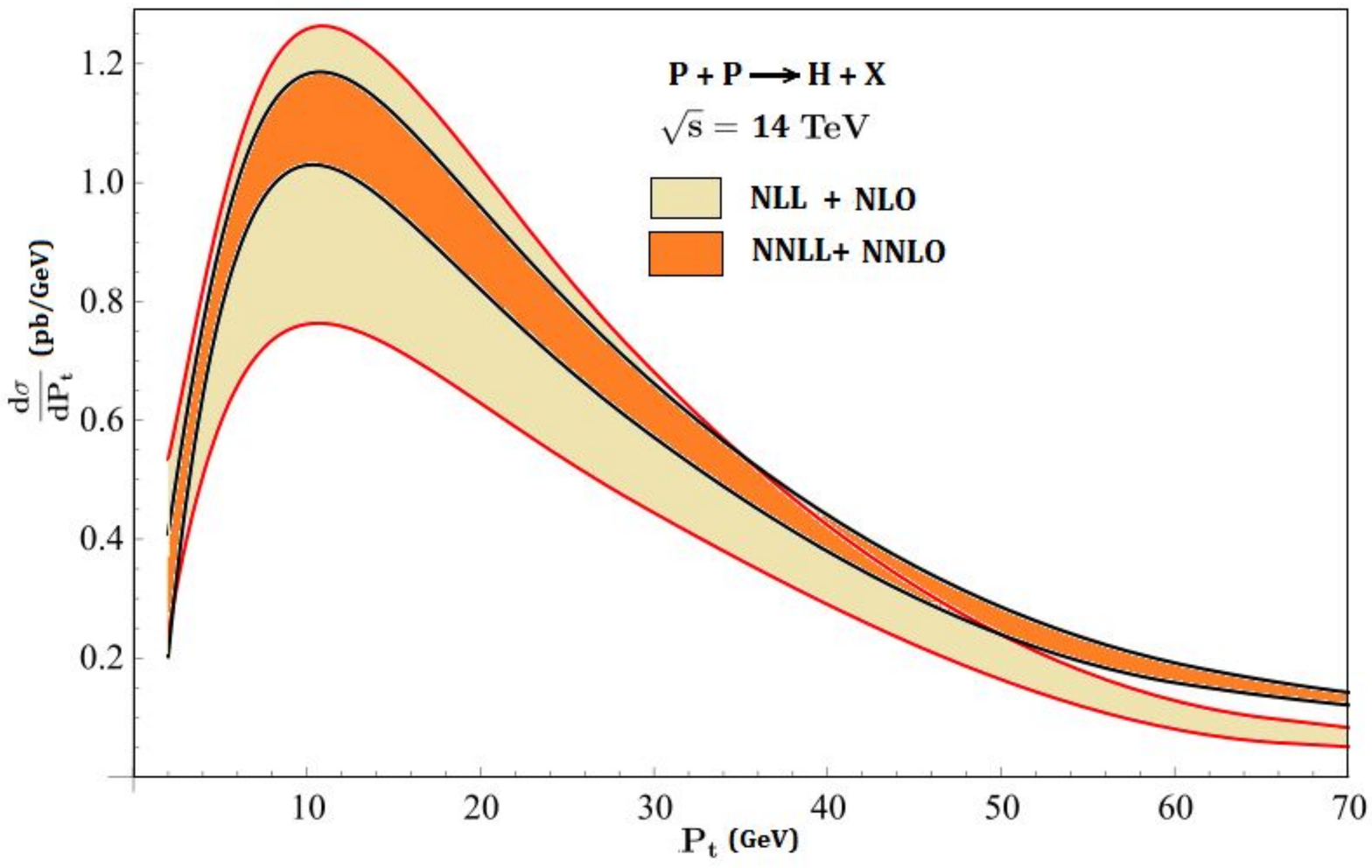}
    \caption{Cross section at 14 TeV}
    \label{14TeV}
\end{figure}
\begin{figure}
      \centering
			\includegraphics[width=5in]{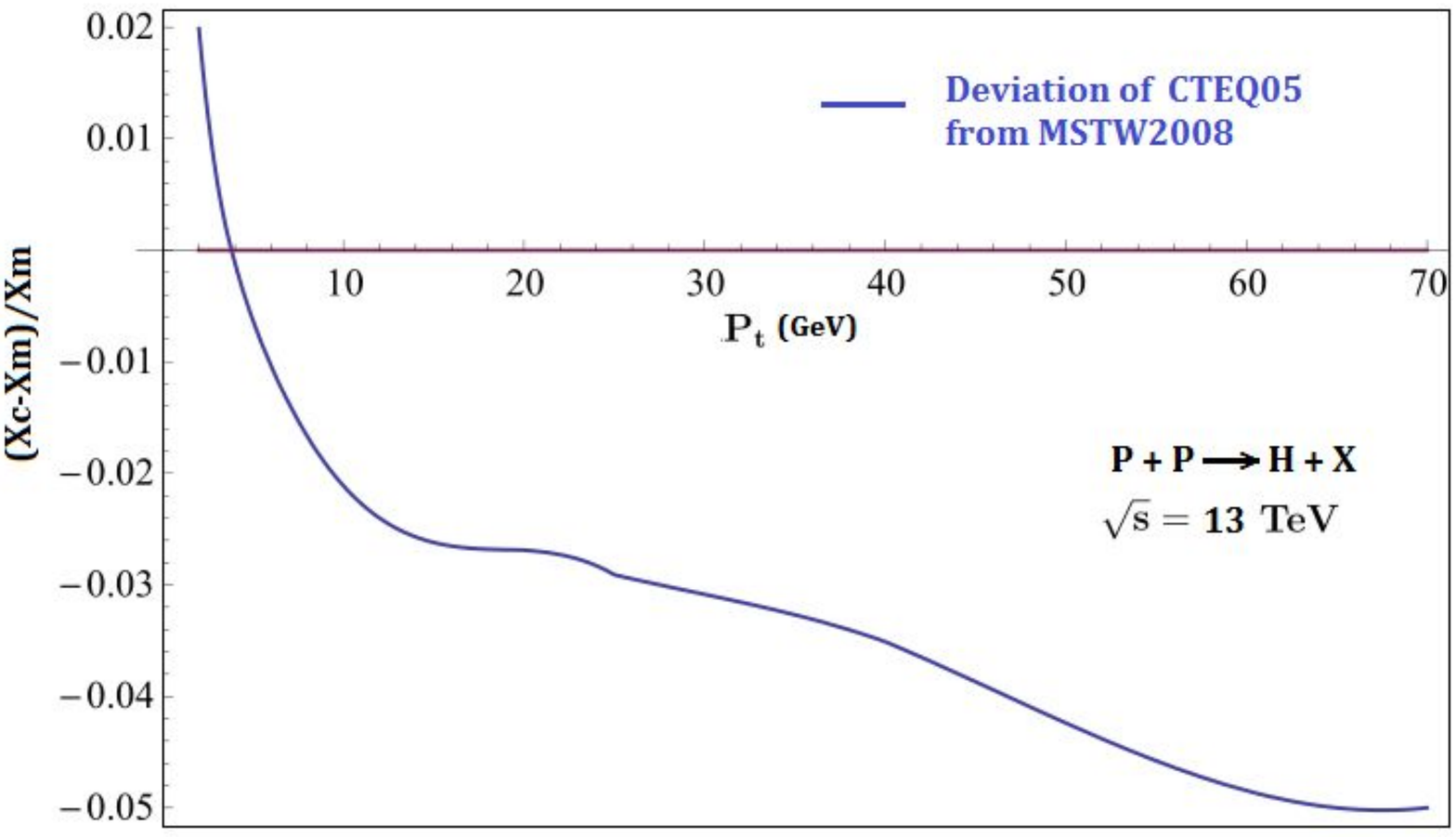}
    \caption{Comparison between pdf sets MSTW2008 and CTEQ05. The y axis plots the  fraction of the difference between the central values of the cross section using the two sets}
    \label{cteq}
  \end{figure}
\begin{figure}
      \centering
			\includegraphics[width=5in]{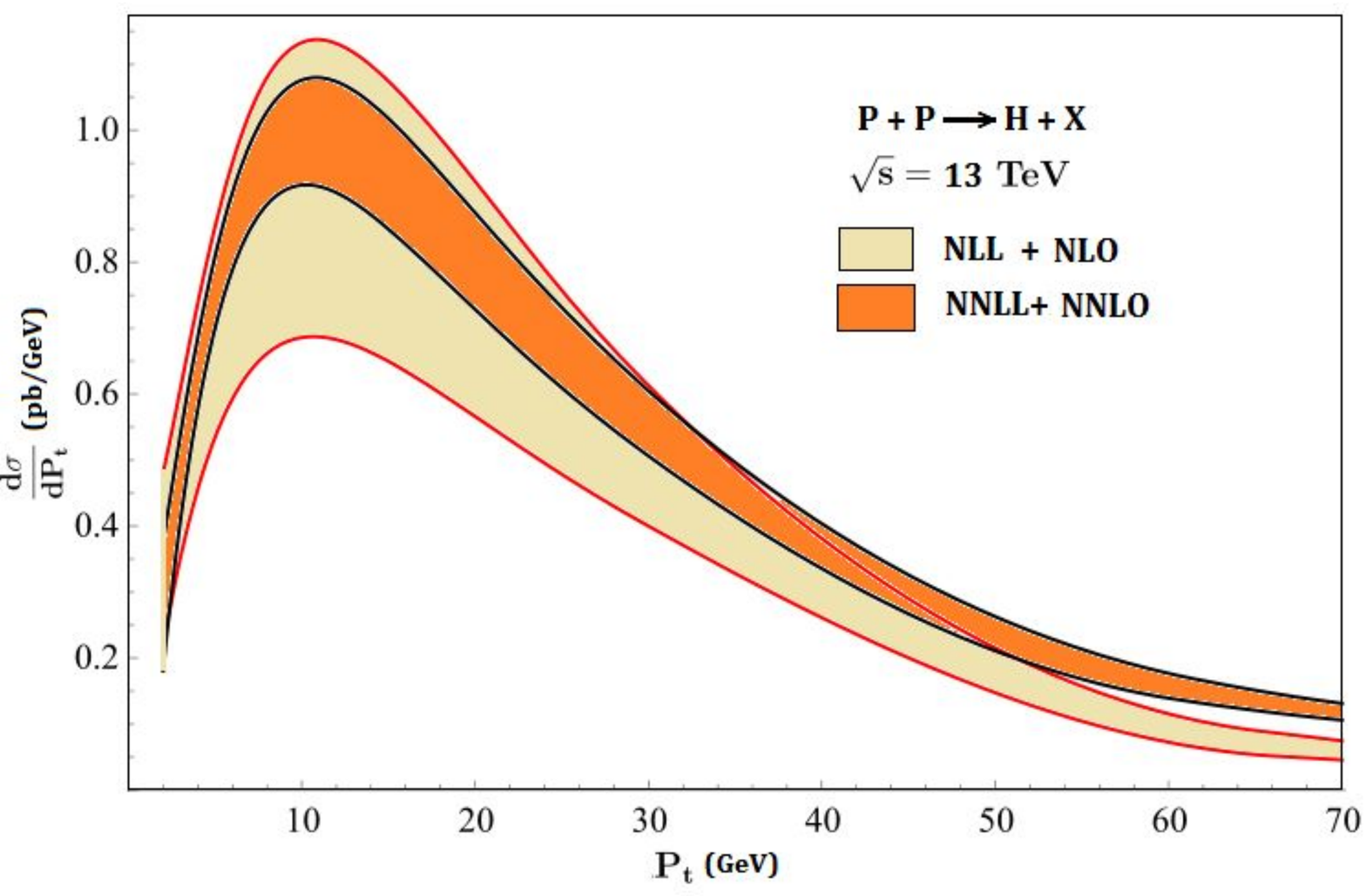}
    \caption{Cross section at 13 TeV}
    \label{13TeV}
  \end{figure}

This variation is accomplished by adjusting  the parameter $L$ in each profile 
which  sets the scale for the low scale matching of the jet and soft function  that dominates
 the perturbative errors.
 In keeping with the canonical recipe adopted by the community we vary  $L$ by  a factor of two about its central value($1/b_0$). The corresponding modification in the profiles for either $\mu$ or $\nu$ is shown in the 
Fig.\ref{var1}.
 When we vary these scales we must keep in  mind the restriction that the argument of each large log is varied at most by a factor of two. This gives us a  constraint $0.5 \leq \mu/\nu \leq 2$ since there are logs of the form $\Log(\mu/\nu)$ in the exponent. In all, this provides us with six different possibilities for each of the three profiles (Fig. \ref{pro}). We do a similar analysis for the end point region of each profile by varying the scale $H$ (Fig.\ref{var2}). Since this region of transverse momentum is dominated by the fixed order cross section, the variation in $H$ amounts to a fixed order scale variation.  We choose to keep the largest error band generated due to these variations.
The plots which combine all the effects above, are shown in Fig.\ref{8TeV} and Fig.\ref{14TeV} for $S=8$ and $14~$TeV. We use the MSTW 2008 pdf at NNLO  \cite{Martin:2009iq} and the value $\alpha_s(M_z) =0.1184$. \\
We also consider how the cross section is typically affected by using other pdf sets. In particular we evaluate the cross section at 13 TeV with the CTEQ5 pdf.  The deviation form the MSTW2008 result is of the order of a few percent as shown in Fig.\ref{cteq}. We include these errors as well in the final plot for the cross section at 13 TeV Fig.\ref{13TeV}.

\section{Comparison to Previous Results}

	\begin{figure}
      \centering
			\includegraphics[width=5in]{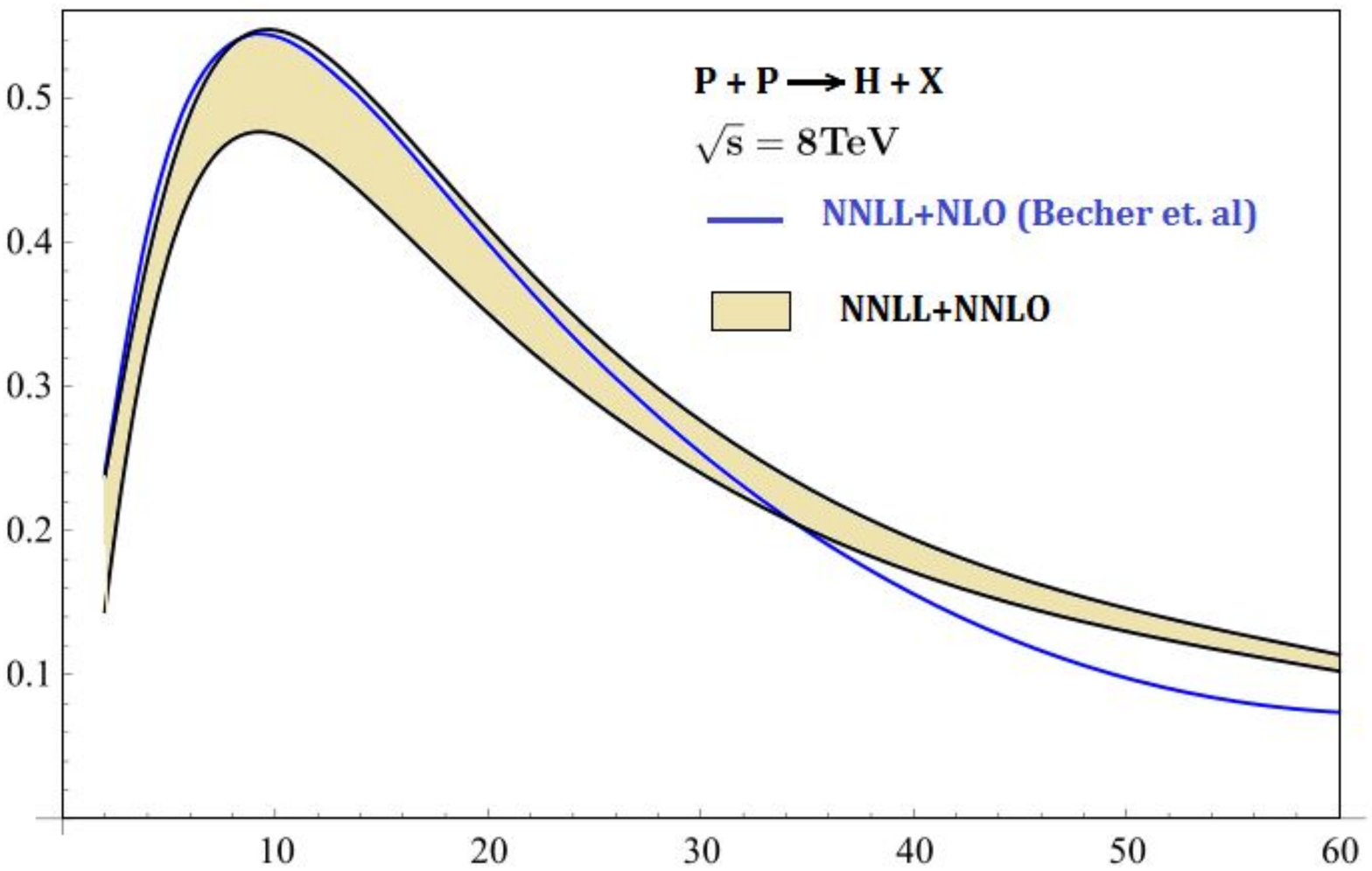}
    \caption{Comparison with Becher et. al.}
    \label{comp_n}
  \end{figure}
	
	\begin{figure}
      \centering
			\includegraphics[width=5in]{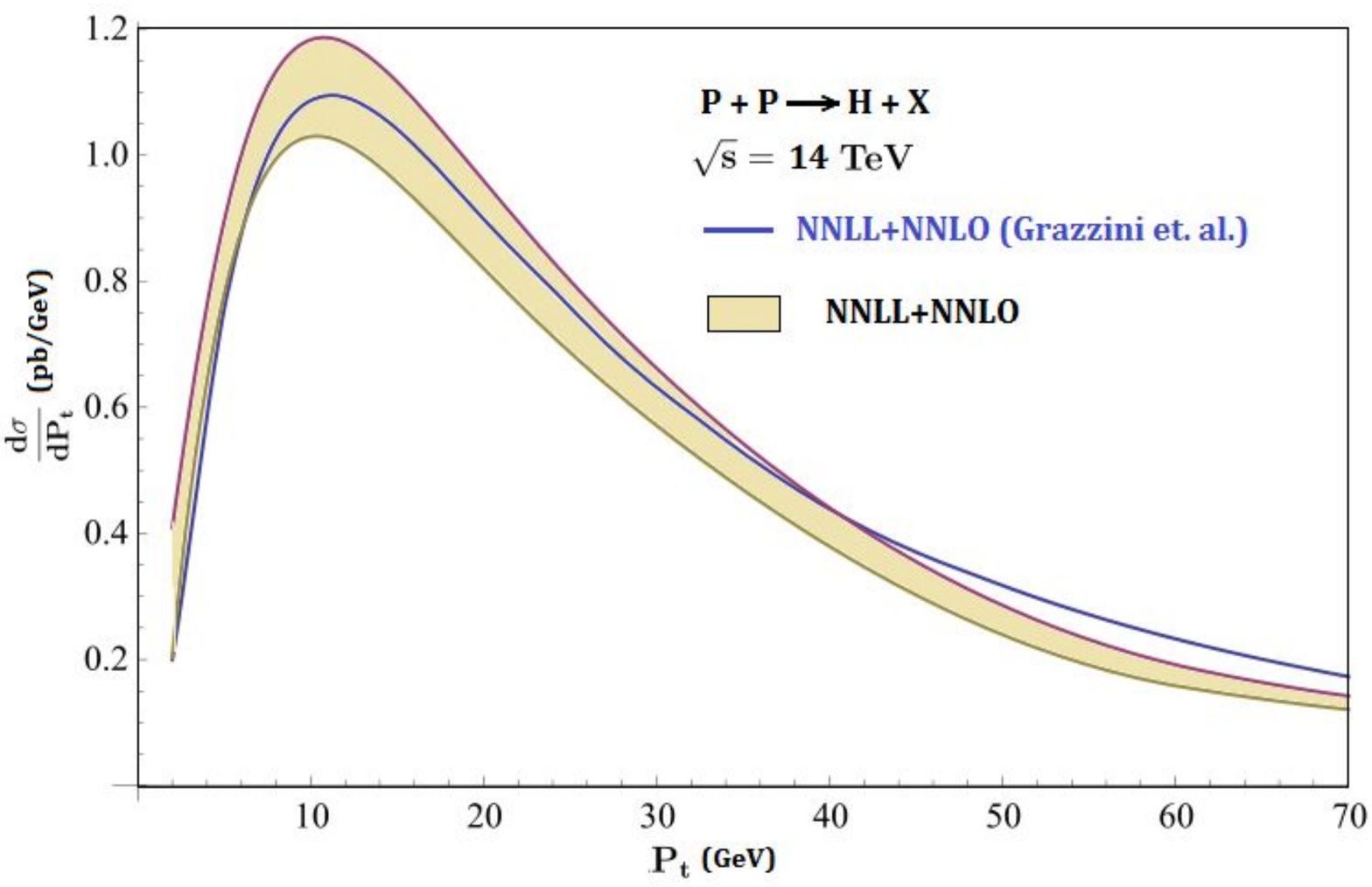}
    \caption{Comparison with CSS formalism}
    \label{comp}
  \end{figure}

Let us now compare our results  with the previous results in the literature which include the $NNLL$
resummation.  We will compare only with those papers
 which show plots which include all of the corrections considered here, i.e. \cite{Grazzini_Figures,higgs_becher}.
We agree formally with the results of these papers.  Of course formal agreement is not necessarily
the relevant issue. Formal agreement means that the coefficients of the  logs in the
resummed result (as well as the non-singular pieces) agree. However, it
does not mean that the argument of the logs are the same. Indeed for sufficiently different
arguments the results can differ by an amount which is well outside any ``reasonable'' theoretical
error estimation. 

In the work  \cite{higgs_becher} the authors  calculate the spectrum up to $60~GeV$ and 
  match onto the NLO result.   Terms of order $\alpha^2$  are not included which is consistent as long
as one presumes that $\frac{\alpha}{\pi} \LN(60^2/mh^2)\sim 1$. That is, as long as the logs still dominate all the way up to $p_\perp=60~GeV$,  then
keeping the terms of order $\alpha^2$ would not improve the accuracy of the calculation.
The results in \cite{higgs_becher}  dont track the rapidity scale. Meaning that an implicit
fixed matching scale has been chosen. In contrast to \cite{higgs_becher}, the NNLL error band in our analysis lies within the error band of NLL.  
The reason is mainly two fold. They have no rapidity scale to vary and they have no errors due to non-perturbative
corrections. This elucidates the importance of including the scale $\nu$ in error analysis for a better estimation of accuracy at each order in resummation. Further, the authors claim the existence of a non-perturbatively generated hard scale 
on the order of $8~GeV$, which they state cuts off the non-perturbative corrections and therefore there results
are valid down to vanishing $p_\perp$.
The underlying justification  is based on the paper
\cite{Parisi:1979se} where it was argued that since in QED the probability to emit a photon with $p_\perp\sim 0$ is vanishes faster then
any value of $Q^2$, the only contribution to the rate must come from two photons whose total $p_\perp$ approximately vanishes but
whose individual $p_\perp$ is large. The authors of \cite{Parisi:1979se} then conjecture that the same reasoning should hold
in the non-Abelian case. A recent study of the non-peturbative contributions to the TMPDF can be found in
\cite{collins} and we invite the reader to consult this paper for further information on the validity of these claims.
%
Fig. \ref{comp_n} shows a comparison with their central value at NNLL+NLO. For larger values of $p_\perp$, the central value curve is below our estimates since the matching is done with fixed order NLO.

Grazzini et. al.  \cite{Grazzini_Figures} use the CSS formalism and plot the result out to $120$ GeV, matching onto the fixed order result
at NNLO.  They do include an additional scale they call the resummation scale $Q$, in addition to the factorization scale of the PDFs and the renormalization scale of $\alpha$. The variation of $Q$ variation mixes what we would call $\mu$ and $\nu$ radiation. This is discussed in detail in the appendix. 
To turn off the resummation in the exponent they make the replacement
\beq
Log(Q^2 b^2/b_0) \rightarrow Log(Q^2 b^2/b_0+1)
\eeq 
this has  the effect of killing the resummation at asymptotically small $b$ and insuring that the total inclusive cross section in
reproduced. However, in the range $M_h/p_\perp\sim 1$ the resummation 
is not shut off rapidly enough. 
Indeed, our results disagree with \cite{Grazzini_Figures} for the central values by an amount
which is larger then the theory error bars in the tail region as can be seen in Fig. \ref{comp}. 
The reasons for this overshoot, as  discussed above, is that if one does not use a
profile to shut off the resummation beyond the transition region, then  the scale of the log in the large $p_\perp$ region is such that the singular and non-singular terms in the fixed order cross section no longer cancel as they do in the fixed order result.  This effect is a recurring theme in the
literature \cite{bsg,thrust,BS}. The authors of \cite{Grazzini_Figures} also include the two loop matching to the hard function  and show that its effects are  at the 
 one percent level.
Moreover, without the other pieces of the calculation of this order, 4-loop cusp, 3 loop non-cusp, two loop soft and collinear matching, this
inclusion does not increase the accuracy of the calculation. Regarding the scale variation, the authors vary three distinct scales as
was performed in this paper. The variations mix rapidity and renormalization group scale dependence (see appendix) and one could argue that
this is perhaps not as clean as varying $\mu$ and $\nu$ independently. 
But given that the scale variation technique is a blunt instrument to estimate perturbative uncertainty, it is not clear that distinguishing logs is quantitatively relevant, as long as all resummed logs are probed in the variation. The uncertainty bands in \cite{Grazzini_Figures} are very close to those presented here. In the appendix we show how the
work of Grazzini et. al. can be parsed in terms of the RRG.

Finally, a recent paper \cite{Echevarria:2015uaa} working within the TMDPDF formalism of \cite{idilbi_pt_NNLL, idilbi_TMDs, idilbi_TMDs_Drell_Yan} published a cross-section for the Higgs transverse momentum distribution at NNLL, with no fixed order matching. The error bands are very small, with no overlap between the NLL and the NNLL resummation, which the authors noted as a sign that they had under-estimated the perturbative uncertainty. The bands were generated by varying the resummation scales generated by the $\mu$ evolution of the TMDPDFs, after the resummation of the rapidity logs, which followed the procedure of \cite{Collins:2011zzd}. The rapidity resummation scales were taken as fixed quantities. Thus no attempt was made to directly gauge the \emph{perturbative} uncertainty in the rapidity resummation, that is, the perturbative uncertainty in the Collins-Soper kernel, whose exponentiation also resums the rapidity logarithms. Interestingly, variation of the non-perturbative models for the Collins-Soper kernel produced a much larger impact on the cross-section, see Fig. 4 of \cite{Echevarria:2015uaa}. Since these non-perturbative models are explicitly exponentiated with the perturbative Collins-Soper kernel, in our view, this variation of the non-perturbative model parameters then may not be an estimate of the non-perturbative physics at all, but a more indirect gauge of the perturbative truncation of the Collins-Soper kernel and the rapidity resummation. Separating out the effects of the perturbative uncertainty in the Collins-Soper kernel, which has not been studied in the literature, will prove important in isolating the non-perturbative physics of the TMDPDFs.

\section{Conclusion}

The next LHC runs promises to shed more light on the nature of the Higgs boson.
Here we have attempted to give the most accurate result possible for the transverse Higgs spectrum using the currently available theoretical calculations for matching coefficients, anomalous dimensions as well as  the fixed order cross section. We resum the large logs in the low $p_\perp$ region to NNLL and match this result to the fixed order cross section for high values of transverse momentum. To maintain accuracy over the complete range of the transverse momentum, we have utilized profiles in both the virtuality ($\mu$) and rapidity ($\nu$)  factorization scales  which
 smoothly turns off the resummation and matches onto the NNLO fixed order cross section at large values of $p_\perp$.
This procedure prevents the problem of enhancing the fixed order result.
 We discuss the sources of error and ways to effectively estimate them using possible  variations in the rapidity and renormalization scales for the profiles introduced. While we get a good agreement with previous results in the low $p_\perp$ region, we get a lower estimate for the cross section in the transition and tail region. 
 
 The code which generated the plots in this paper is available upon request.

\section{Acknowledgements}
This work is supported by DOE contracts DOE-ER-40682-143 and DEACO2-C6H03000. D.N. is also supported by a MIT Pappalardo fellowship, and DOE contracts DE-SC00012567 and DE-SC0011090.
  D.N. also wishes to thank Simone Marzani for discussions of the literature.

\appendix
\section{RRG resummation}
In this appendix, we review  the RRG formalism for the purpose of explicitly comparing 
to the method used in \cite{Grazzini_I, Florian, Bozzi:2005wk}. After integrating over the rapidity of the Higgs, the differential cross section is given by
\begin{align}\label{fact_for_xsec}
\frac{d\sigma}{dp_t^2}&=\sigma_0\int \frac{dx_1}{x_1} \frac{dx_2}{x_2}\int d^2\vec{b} e^{i\vec{b}\cdot \vec{p}_t}\delta(m_h-Sx_1x_2)H(m_h,\mu)\nonumber\\
&\qquad\qquad f_{n\perp}\Big(x_1,\vec{b},\mu,\nu\Big)f_{\nbar\perp}\Big(x_2,b,\mu,\nu\Big)S\Big(\vec{b},\mu,\nu\Big)\,,
\end{align}
where we have written out the explicit momentum fraction dependence. Since we are focusing on the resummation, we will often suppress the other arguments of the renormalized functions. 
The soft and beam functions have the $\mu$-anomalous dimension:
\begin{align}
\mu\frac{d}{d\mu}\text{ln} S(\nu,\mu)&=\Gamma_{R}[\alpha_s(\mu)]\text{ln}\Big[\frac{\mu}{\nu}\Big]+\gamma_s(\alpha_s(\mu))\,,\\
\mu\frac{d}{d\mu}\text{ln} f_{\perp}(\nu,\mu)&=\frac{1}{2}\Gamma_{R}[\alpha_s(\mu)]\text{ln}\Big[\frac{\nu}{m_h}\Big]+\gamma_{f_{\perp}}(\alpha_s(\mu))\,.
\end{align}

Using the fact that
the RG and RRG  variations of the amplitude must commute, we can write the rapidity anomalous dimension as 
integral over the $\mu$ anomalous dimensions \cite{RRG1,RRG2} and a constant piece
\begin{align}
\nu\frac{d}{d\nu}\text{ln} S(\nu,\mu)&=\int_{\binv}^{\mu}\frac{dq}{q}\Gamma_{R}[\alpha_s(q)]+\gamma_R\Big[\alpha_s(\binv)\Big]\,,\\
\nu\frac{d}{d\nu}\text{ln} f_{\perp}(\nu,\mu)&=-\frac{1}{2}\int_{\binv}^{\mu}\frac{dq}{q}\Gamma_{R}[\alpha_s(q)]-\frac{1}{2}\gamma_R\Big[\alpha_s(\binv)\Big]\,.
\end{align}
Writing things in this way allows for there to be large $\mu$ dependent logs which may arise depending
upon the choice of $\mu$.  As noted in Sec. \ref{sec:Renorm_Resum}, $\Gamma_R$ is related to the cusp anomalous dimension. The non-cusp piece of the rapidity anomalous dimension $\gamma_R\Big[\alpha_s(\binv)\Big]$ can be extracted from the calculation of the soft function at the renormalization point $\mu=\binv$.  This is a choice of scheme which defines the scale in the Log  which multiplies the cusp anomalous dimension.

We run the beam/soft functions in $\nu$ from $\nu_s$ and $m_h$ to $\nu$ respectively
\begin{align}
S\Big(\vec{b},\mu,\nu\Big)&=U^{RRG}_S\Big(\vec{b},\mu,\frac{\nu}{\nu_s}\Big)S\Big(\vec{b},\mu,\nu_s\Big)\,,\\
f_{n\perp}\Big(x,\vec{b},\mu,\nu\Big)&=U^{RRG}_F\Big(\vec{b},\mu,\frac{\nu}{m_h}\Big)f_{n\perp}\Big(x,\vec{b},\mu,m_h\Big)\,.
\end{align}
Note that we have explicitly chosen $\nu_B=m_h$ in the TMDPDF. The full RRG factor has the rapidity factorization scale $\nu$ cancel in the exponent, and gives:
\begin{align}
U^{RRG}_{SF^2}\Big(\vec{b},\mu,\frac{m_h}{\nu_s}\Big)=\text{Exp}\Bigg[\text{ln}\Big(\frac{m_h}{\nu_s}\Big)\Big\{\int_{\binv}^{\mu}\frac{dq}{q}\Gamma_{R}[\alpha_s(q)]+\gamma_R\Big[\alpha_s(\binv)\Big]\Big\}\Bigg]\,.
\end{align}
We also need the $\mu$ evolution of the TMDPDFs at $\nu=m_h$ and the soft function at $\nu=\nu_s$. With these scale choices, the TMDPDF has no double log-$\mu$ evolution, so one obtains:
\begin{align}\label{eq:low_scale_mu_evo}
f_{\perp}(x,\vec{b},\mu,\nu=m_h)&=\text{Exp}\Bigg[\int_{\mu_i}^{\mu}\frac{dq}{q}\gamma_{f_\perp}\big[\alpha_s(q)\big]\Bigg]f_{\perp}(x,\vec{b},\mu_i,\nu=m_h)\,,\nonumber\\
S(\vec{b},\mu,\nu=\nu_s)&=\text{Exp}\Bigg[\int_{\mu_i}^{\mu}\frac{dq}{q}\Big\{\LN\Big(\frac{\nu_s}{q}\Big)\Gamma_R[\alpha_s(q)]+\gamma_s[\alpha_s(q)]\Big\}\Bigg]S(\vec{b},\mu_i,\nu=\nu_s)\,.
\end{align}
Combining these pieces together, we achieve the full resummation kernel for the cross-section:
\begin{align}
f_{n\perp}(\mu,\nu)f_{\bar n\perp}(\mu,\nu)S(\mu,\nu)&=R\Big(\mu,\mu_i,\nu_s,m_h,b\Big)f_{n\perp}(\mu_i,m_h)f_{\bar n\perp}(\mu_i,m_h)S(\mu_i,\nu_s)\,\\
\label{eq:finall_RRG_RG}R\Big(\mu,\mu_i,\nu_s,m_h,b\Big)&=\text{Exp}\Bigg[\int_{\mu_i}^{\mu}\frac{dq}{q}\Big\{\text{ln}\Big[\frac{m_h}{q}\Big]\Gamma_R[\alpha_s(q)]+\gamma_s[\alpha_s(q)]+2\gamma_{f_\perp}[\alpha_s(q)]\Big\}\nonumber\\
&\qquad+\text{ln}\Big[\frac{m_h}{\nu_s}\Big]\Big\{\int_{\binv}^{\mu_i}\frac{dq}{q}\Gamma_{R}[\alpha_s(q)]+\gamma_R\Big[\alpha_s(\binv)\Big]\Big\}\Bigg]
\end{align}
From the form of \eqref{eq:finall_RRG_RG}, we can directly see the variation of the independent resummation scales $\mu,\mu_i,\nu_s$ probe different exponentiated logs, thus giving estimates of  the subleading terms in the perturbative expansions. The double logarithmic terms associated with $\mu_i$ cancel manifestly in the exponent\footnote{Though depending on how one handles the integration over the running coupling, this cancellation could be incomplete. An incomplete cancellation could be used to give another handle on estimating the uncertainty in the double logarithmic terms.}, and thus the $\mu_i$ variation is estimating the subleading terms associated with $\gamma_{s,f_\perp}$. We can eliminate large logs of the impact parameter from the beam and soft function matrix elements by choosing the scales $\mu_i,\nu_s$ appropriately. Four potential schemes for canonical scale choices are:
\begin{itemize}
\item[A]: $\mu_i=\nu_s=\binv$
\item[B]: $\mu_i=p_T$, $\nu_s=\binv$
\item[C]: $\mu_i=\binv$, $\nu_s=p_T$
\item[D]: $\mu_i=\nu_s=p_T$
\end{itemize} 
As noted in \cite{sterman_qcdvscet}, making scale choices in either momentum space or conjugate space can have a sizable numerical impact on the cross-section, as well as how accurately the resummation captures the higher order logs claimed by the resummation. Scale setting in momentum space, residual logs can exist in the plus distributions of the low scale theory that become apparent only after integrating the resummation kernel against the matrix elements, in this case, the TMDPDF and the soft function. Scale setting in conjugate space does not suffer from this ambiguity. Specifically, for the canonical choice $\nu_s=\mu_i=\binv$, $R$ becomes:
\begin{align}\label{RRG_RG_Final_bspace}
R\Big(\mu,m_h,b\Big)&=\text{Exp}\Bigg[\int_{\binv}^{\mu}\frac{dq}{q}\Big\{\text{ln}\Big[\frac{m_h}{q}\Big]\Gamma_R[\alpha_s(q)]+\gamma_s[\alpha_s(q)]+2\gamma_{f_\perp}[\alpha_s(q)]\Big\}\nonumber\\
&\qquad\qquad+\text{ln}\Big[m_h be^{\gamma_E}\Big]\gamma_R[\alpha_s\big(\binv\big)]\Bigg]\,.
\end{align}

\subsection{Comparision to Grazzini et al.}
We now write out the resummation formalism used in \cite{Grazzini_Figures}, as derived in \cite{Grazzini_I, Florian, Bozzi:2005wk}, suppressing the flavor sum in the PDFs:
\begin{align}
\frac{d\sigma}{dp_t^2}&=\int \frac{dx_1}{x_1} \frac{dx_2}{x_2}\int \frac{d^2\vec{b}}{4\pi} e^{i\vec{b}\cdot \vec{p}_t} f(x_1,\mu_F)f(x_2,\mu_F){\mathcal W}\Bigg(b,m_h,x_1x_2,\mu_R,\mu_F\Bigg)
\end{align}
Then one switches to moment space by integrating $\frac{1}{z}=x_1x_2$ for ${\mathcal W}$:
\begin{align}
{\mathcal W}_N\Bigg(b,m_h,\mu_R,\mu_F\Bigg)&=\int_0^1 dz \,z^{N-1}{\mathcal W}\Bigg(b,m_h,\frac{1}{z},\mu_R,\mu_F\Bigg)
\end{align}
The authors write this as 
\begin{align}\label{grazzini_hard_func_resum}
{\mathcal W}_N\Bigg(b,m_h,\mu_R,\mu_F\Bigg)&={\mathcal H}_N\Bigg(m_h,\mu_R,\mu_F;Q\Bigg)\text{Exp}\Bigg[{\mathcal G}_N\Bigg(b,m_h,\mu_R,Q\Bigg)\Bigg].
\end{align}
Note that the factorization scale from the PDFs is in the function ${\mathcal H}$. Then the form of the resummed exponent is:
\begin{align}\label{g_resum}
{\mathcal G}_N\Bigg(b,m_h,\mu_R,Q\Bigg)&=-\int_{b^2 e^{2\gamma_E}}^{Q^2}\frac{dq^2}{q^2}\Bigg\{A\Big[\alpha_s(q)\Big]\text{ln}\Big(\frac{m_h^2}{q^2}\Big)+B_N\Big[\alpha_s(q)\Big]\Bigg\}
\end{align}
Note that $B_N$ has PDF running in it since it depends on the moment $N$. The function ${\mathcal H}_N$ is claimed to have no large logarithms, and can be perturbatively calculated. To one loop:
\begin{align}
{\mathcal H}_N\Bigg(m_H,\mu_R,\mu_F;Q\Bigg)&=1+\frac{\alpha_s}{\pi}\Bigg(H^{(1)}+2C_N^{(1)}-p\beta_0\ell_R+2\gamma_N\ell_F-(\frac{1}{2}A^{(1)}\ell_Q+B^{(1)}+2\gamma_N^{(1)})\ell_Q\Bigg)
\end{align}
Where:
\begin{align}
\ell_R=\text{ln}\frac{m_h^2}{\mu_R^2} & & \ell_F=\text{ln}\frac{m_h^2}{\mu_F^2} & &\ell_Q=\text{ln}\frac{m_h^2}{Q^2}
\end{align}
And:
\begin{align}
A^{(1)}&=C_A\\
B^{(1)}&=-\frac{1}{6}(11C_A-2N_f)\\
\gamma_N^{(1)}&=\int_0^1dz\, z^{N-1}P^{(1)}(z)
\end{align}
$C_N$ is related to the order $\epsilon$ pieces of the DGLAP splitting kernels. $H^{(1)}$ is the hard matching, but also contain terms in the $\delta(1-z)$ pieces of the TMDPDF. Note that $\mu_R\sim\mu_F\sim Q\sim m_h$, so that the PDFs are run from $\Lambda_{QCD}$ to the high scale, implicitly.

By choosing the appropriate scales, we can connect the resummed formula of \eqref{g_resum} to the resummation formulas of \eqref{eq:finall_RRG_RG} and \eqref{RRG_RG_Final_bspace}. We must also apply the resummation scale prescription of \cite{Bozzi:2005wk}, where all large logs are split:
\begin{align}\label{resummation_scale_prescription}
\text{ln}[m_h\binv]\rightarrow\text{ln}\Big[\frac{m_h}{Q}\Big]+\text{ln}\Big[Q\binv\Big]
\end{align}
and any log of $\text{ln}\Big[\frac{m_h}{Q}\Big]$ is expanded out of the exponent and included in the hard function of \eqref{grazzini_hard_func_resum}. Setting $\mu=Q$ in \eqref{RRG_RG_Final_bspace}, this gives:
\begin{align}
\label{R}
R(Q,m_h,b)&=\text{Exp}\Bigg[\int_{\binv}^{Q}\frac{dq}{q}\Big\{\text{ln}\Big[\frac{m_h}{q}\Big]\Gamma_R[\alpha_s(q)]+\gamma_s[\alpha_s(q)]+2\gamma_{f_\perp}[\alpha_s(q)]+\gamma_R[\alpha_s\big(\binv\big)]\Big\}\nonumber\\
&\qquad\qquad+\text{ln}\Big[\frac{m_h}{Q}\Big]\gamma_R[\alpha_s\big(\binv\big)]\Bigg],
\end{align}
choosing $Q=\mu$ this result is identical to (\ref{RRG_RG_Final_bspace}).

The hard function of Eq. \eqref{grazzini_hard_func_resum} is explicitly given to two loops in \cite{Bozzi:2005wk} , and one can verify the factor $\text{ln}\Big[\frac{m_h}{Q}\Big]\gamma_R[\alpha_s\big(\binv\big)]$ in (\ref{R}) is expanded out of the exponent and then included in the $B^{(2)}$ term of the hard matching. Indeed, from the form of the hard function given in \cite{Bozzi:2005wk}, one can see the resummation scale $Q$ probes both rapidity logs and RG logs since $B^{(2)}$ includes both types of anomalous dimensions. This explains the similar sized perturbative error estimation of this paper, and that of \cite{Grazzini_Figures}. 

\end{document}